# Evaluation du score mitotique des carcinomes mammaires infiltrants sur lame numérique : développement et apport d'un algorithme de détection de mitoses

Loris Guichard1, Clara Simmat2, Margot Dupeux1, Stéphane Sockeel2, Nicolas Pozin2,

Magali Lacroix-Triki3, Catherine Miquel4, Marie Sockeel2, Sophie Prévot1

1 Hôpital Bicêtre (AP-HP), Le Kremlin-Bicêtre, 2 Primaa, Paris, 3 Institut Gustave Roussy, Villejuif, 4 Hôpital Saint-Louis (AP-HP), Paris

*Note préliminaire : le travail exposé ci-dessous correspond à la phase de 3 de l'étude entreprise dans le cadre du travail de thèse réalisé précédemment intitulé « Evaluation du score mitotique des carcinomes mammaires infiltrants : développement et apport d'un algorithme de détection de mitoses », soutenue le 12 octobre 2022 et dont les résultats ne figuraient pas au sein du manuscrit.*

# Introduction

Le cancer du sein est un problème majeur de santé publique. C'est le cancer le plus fréquent tous sexes confondus et la première cause de mortalité par cancer chez la femme à travers le monde. En 2020, plus de 2,2 millions de cas ont été recensés et on estime qu'il a été la cause d'environ 685 000 décès (1). Toutefois c'est également un cancer de bon pronostic. Lorsqu'il est pris en charge à temps, le taux de survie à 5 ans est de 90% dans les pays développés.

Pour tout cancer du sein nouvellement diagnostiqué, Le grade d'Elston-Ellis est un facteur pronostic majeur. Il s'agit d'un grade en trois catégories déterminées elles même par l'évaluation de trois paramètres que sont le pourcentage de formations glandulaires présentes dans la tumeur, le degré d'atypie des noyaux et le compte mitotique pour une surface tumorale donnée (2). Ce grade conditionne en partie la prise en charge des patientes. Pour environ 30% des patientes atteintes d'un cancer du sein, la décision d'un traitement par chimiothérapie ou hormonothérapie dépendra directement de l'évaluation de ce grade (3).

Le score mitotique est le reflet de l'activité proliférative de la tumeur. Il est le seul des trois items du grade d'Elston-Ellis à avoir une valeur pronostique indépendante quel que soit la taille tumorale ou le statut ganglionnaire (4). Ce score mitotique est évalué en deux étapes. Tout d'abord, la détermination de la zone de *hotspot* correspondant à la zone considérée par le pathologiste comme la plus mitotique généralement située en périphérie de la tumeur, puis le compte mitotique dans cette zone sur une surface donnée (généralement de 2 ou 3mm²) correspondant à 10 champs à fort grossissement. La sélection de la zone et l'évaluation des figures mitotiques posent des problèmes de reproductibilité inter-observateur et intra-observateur (5,6).

Depuis la fin des année 50 et l'introduction du grade par Scarff, Bloom and Richardson, les critères morphologiques d'évaluation des mitoses et la méthode de comptage ont régulièrement évolué pour essayer d'améliorer l'objectivité des pathologistes dans leur évaluation (7–9).

Aujourd'hui l'approche du pathologiste dans l'évaluation du cancer du sein et en particulier dans l'évaluation des mitoses est à même d'être modifiée avec le développement de la pathologie numérique (10,11).

Sur un microscope, la vis micrométrique permet de faire varier le plan focal en navigant dans toute l'épaisseur de la coupe ce qui facilite la détection des mitoses.

La lame numérique correspond à une représentation de la lame de verre sur un seul plan focal sans possibilité de naviguer dans l'épaisseur de la coupe. L'équipe de Nottingham a récemment montré que les détails morphologiques des mitoses étaient moins nets sur lame numérique pouvant entraîner un risque de sous-estimation du score mitotique. Ils ont également formulé des recommandations pour identifier plus précisément les mitoses sur lame numérique (12,13).

Certaines approches telles que la numérisation multi-plan pourraient être envisagées pour résoudre ce problème de détection des mitoses mais ceci au prix d'une augmentation drastique du temps de numérisation et du poids des lames numérisées ainsi que d'un manque d'instantanéité de lecture rendant son utilisation en routine assez fastidieuse (14).

Une des pistes avancées pour essayer de résoudre le problème de l'évaluation des mitoses sur lame numérique est l'utilisation d'outils d'intelligence artificielle et en particulier les approches d'IA basées sur les réseaux de neurones convolutifs. Ces modèles entraînés le plus souvent à partir de bases de données labellisées sont constitués de multiples neurones artificiels interconnectés qui échangent des informations permettant de décomposer l'analyse des images pour résoudre des tâches de détection ou de classification variées (15,16).

Depuis 2012, de nombreux challenges ont été organisés pour développer des algorithmes de détection de mitoses. Ces challenges permettent aux développeurs de confronter leurs méthodes mais n'ont pas forcément pour finalité l'intégration de l'outil dans le flux de travail du pathologiste, les résultats portant essentiellement sur les métriques de performance des algorithmes. De plus, la qualité des lames et les paramètres de numérisation ne sont pas toujours représentatifs de la routine diagnostique ce qui peut freiner la transition vers une utilisation des modèles même les plus performants dans la vraie vie (17–19).

Peu d'études se sont penchées sur l'apport concret de ces outils pour le pathologistes dans une démarche d'utilisation en routine diagnostique.

Dans l'étude de Veta *et al.*, (20) portant sur l'évaluation de 100 cas de carcinomes infiltrants par trois pathologistes sur 2 mm² de surface, les discordances entre algorithme et pathologistes pour la détermination du score mitotique étaient supérieures à celles observée entres les pathologistes eux mêmes.

Dans l'étude de Balkenhol *et al.* (21) portant sur 90 cas de carcinomes infiltrants du sein, la reproductibilité de deux pathologistes dans l'évaluation du score mitotique était améliorée sur zone de *hotspot* de 2mm² proposée par algorithme sur lame numérique par rapport à une évaluation sur lame de verre. Cependant dans l'étude le compte mitotique sur lame numérique était réalisé strictement dans la même zone proposée par l'algorithme par les deux observateurs ce qui diffère de la routine diagnostique où les pathologistes analysent la totalité de la surface tumorale pour choisir la zone de *hotspot*, laquelle peut différer d'un pathologiste à l'autre.

Dans l'étude de Pantanowitz *et al.,* (22) vingt-quatre pathologistes de niveau d'expertise variés ont compté les mitoses à deux reprises avec ou sans assistance d'un algorithme sur 140 champs de 0,2 mm² avec une période de wash-out de quatre semaines. La précision et la sensibilité des pathologistes étaient améliorées par l'utilisation de l'algorithme. Néanmoins cette étude portait uniquement sur la détection des mitoses sans évaluation du score mitotique et la reproductibilité des pathologistes n'a pas été évaluée. L'impact potentiel de l'utilisation de l'algorithme sur le score mitotique final n'a donc pas été exploré.

Le but de notre étude était d'évaluer l'apport d'un algorithme de détection de mitoses pour le pathologiste lors de l'évaluation du score mitotique des carcinomes mammaires infiltrants sur lame entière numérisée en condition de routine. Pour se faire, la reproductibilité inter-observateur du compte et du score mitotique de deux pathologistes de niveau différents ont été évaluées avec et sans utilisation de l'algorithme entre eux et avec un consensus de trois pathologistes experts.

## Matériel et méthodes

### Préparation du matériel

Pour cette étude, 50 cas de carcinomes infiltrants du sein (25 biopsies et 25 pièces opératoires) pris en charge dans le service d'anatomopathologie de l'hôpital Bicêtre (AP-HP) ont été sélectionnés de façon aléatoire entre Février 2020 et Juin 2021. Pour chaque cas, une lame représentative de la tumeur a été sélectionnée par un médecin pathologiste spécialisé en pathologie mammaire du service de Bicêtre (SP).

Le **tableau 1** présente les caractéristiques des patients et des tumeurs des 50 cas utilisés pour l'étude.

Les biopsies ont été fixées en formol neutre tamponné à 4% de formaldéhyde pendant au minimum 6h. Les pièces opératoires ont été échantillonnées après fixation en formol tamponné à 4% pendant 24 à 48h. Puis les prélèvements en cassettes ont été déshydratés et imprégnés sur automate Sakura Tissue-Tek® VIP® selon le protocole en vigueur dans le service. L'inclusion en paraffine a été effectuée sur automate Sakura Tissue-Tek® AutoTEC® pour les échantillons de pièces opératoires et manuellement sur platine Sakura Tissue-Tek® pour les biopsies. Les coupes réalisées au microtome Leica ou Microm de 3 µm d'épaisseur ont été étalées sur lame de verre SuperFrost™ puis mises à sécher 30 min à l'étuve à 60 °C et 15 min dans le colorateur. La coloration en Hématoxyline-Eosine-Safran a été effectuée sur automate Leica ST5020®. Les lames ont ensuite été montées sur l'automate Leica CV5030® avec un milieu de montage Pertex® puis séchées au minimum 5 min dans l'automate.

La numérisation des cas a été effectuée sur deux scanners de lames P1000 (pièces opératoires) et P250 (biopsies) de la société 3DHISTECH, tous deux équipés de deux objectifs Plan-Apochromat x20 et x40 et d'une caméra Adimec QUARTZ Q-12A180 d'une définition de 4096 × 3072 pixels - (taille du pixel : 5.5 µm × 5.5 µm) permettant une multiplication numérique x1,6 du grossissement.

Les 50 cas ont été numérisées au format .mrxs avec une définition de 0,24 µm/pixel selon un protocole utilisé en routine diagnostique dans le service à l'objectif x20 avec multiplicateur numérique x1,6.

Pour les besoins de l'étude les cas ont été lus sur moniteur écran 27" haute résolution EIZO FlexScan EV2750.

## Algorithme

L'algorithme développé par la start-up Primaa et utilisé pour ce travail est basé sur deux réseaux de neurones successifs. Un premier réseau de neurone basé sur l'architecture du Retinanet (23) correspond à un extracteur de caractéristiques qui a pour objectif de détecter toutes les cellules pouvant potentiellement correspondre à des mitoses. C'est donc un réseau sensible mais assez peu spécifique. Pour compenser ce manque de spécificité, un second réseau de neurone basé sur le MobileNet (24) est utilisé comme classifieur avec pour objectif de classer toutes les images détectées en première étape soit en mitose soit en "non-mitoses". La **figure 1** présente le fonctionnement général de l'algorithme.

Ces deux réseaux ont été entraînés à partir de deux bases de données. Une première base de données correspondant à 1677 mitoses labellisées sur 32 lames de carcinome infiltrant du sein, prises en charge dans le laboratoire de Bicêtre et différentes de celles utilisées dans le cadre de cette étude. La seconde base de données correspondant à celle utilisée pour le challenge MIDOG2021 (19) disponible en open-access et dont 1483 mitoses ont été utilisées.

Pour permettre au pathologiste d'interagir avec les résultats de l'algorithme, une interface de visualisation nommée Cleo, a été développée par Primaa. Elle permet au pathologiste de visualiser les images détectées par le retinanet et retenues comme mitose par le mobilenet. Les zones de plus forte densité mitotique sont affichées sous la forme de carrés de 1mm². Un onglet latéral permet au pathologiste de visualiser et de repérer la position des mitoses sur la lame sans être obligé de zoomer. Cet onglet permet également de supprimer d'éventuels faux positifs ce qui permet de réajuster les zones de *hotspot*. La **figure 2** présente l'interface d'utilisation utilisée dans le cadre de cette étude.

## Design de l'étude

Pour toutes les phases de l'étude, l'évaluation des cas a été faite en tenant compte des critères récemment publiés par l'équipe de Nottingham pour l'évaluation des mitoses (12,13). Les 50 cas de carcinome infiltrant ont premièrement été vus de façon indépendante par trois pathologistes experts en pathologie mammaire de sorte à établir un consensus sur le score mitotique. Chaque pathologiste expert effectuait le compte mitotique et la détermination du score correspondant en aveugle des deux autres. Pour les cas discordants (17/50), une séance de relecture commune a été effectuée pour établir le consensus.

Secondairement, une lecture des cas avec et sans l'algorithme a été effectuée par deux pathologistes de niveau différents, (interne et assistant-hospitalo-universitaire). Cette lecture était basée sur un schéma croisé ou la moitié des cas ont été lus sans l'algorithme et l'autre moitié avec puis inversion du mode de lecture après une période de *wash-out* de 4 semaines. Pour les deux pathologistes, les cas lus avec et sans l'algorithme en première et en seconde lecture étaient opposés à chacune des deux phases.  Les pathologistes étaient libres d'effectuer leur compte mitotique dans la zone qu'ils estimaient la plus appropriée y compris lors de l'utilisation de l'algorithme. Chaque pathologiste a effectué sa lecture en aveugle de la lecture effectuée par l'autre et par le consensus d'expert.

Pour toutes les phases de l'étude, l'aire de comptage sur la lame numérique et les mitoses retenues par les pathologistes ont été enregistrées directement dans l'interface de lecture. Le compte et le score mitotique de chaque pathologiste pour chacun des 50 cas ont été compilés dans un tableau excel.

## Aspects éthiques et réglementaires

Cette étude a été effectuée conformément aux dispositions du Code de la Santé Publique Français applicables aux recherches n'impliquant pas la personne humaine (Code de la Santé Publique - Article R1121-1 modifié par le Décret n° 2017-884 du 9 mai 2017) et ne relève donc pas de la compétence d'un Comité de Protection des Personnes. Il a obtenu l'avis favorable du Comité d'Expertise pour les Recherches, les Études et les Évaluations dans le domaine de la Santé ainsi que du comité d'éthique de la recherche des Hôpitaux Universitaires Henri-Mondor (n° d'avis 2022-134). Il a fait l'objet d'une soumission auprès de la Commission Nationale de l'Informatique et des Libertés sous la référence 2225201 v 0 et a été déclaré en conformité avec la méthodologie de référence MR-004. Les patientes impliquées ont été informées de la recherche via une note d'information distribuée par courrier postal avec possibilité de s'opposer à l'étude dans un délai de trois semaines à compter de la date d'envoi.

## Analyses statistiques

Les analyses statistiques ont porté sur la reproductibilité inter-observateur du compte et du score mitotique des deux pathologistes lecteurs avec et sans l'algorithme et sur la reproductibilité inter-observateur entre chaque pathologiste lecteur et le consensus d'expert avec et sans l'algorithme.

Pour la reproductibilité du compte mitotique (variable quantitative), le coefficient de corrélation intra-classe (CCI) a été calculé en utilisant un modèle aléatoire à deux facteurs portant sur l'accord absolu entre observateur à partir d'une mesure unique. Les recommandations de Koo et Li de 2016 (25), ont été considérées pour l'interprétation de l'accord : en dessous de 0,50 : faible ; entre 0,50 et 0,74 : moyen ; entre 0,75 et 0,90 : bon ; au-dessus de 0,90 : excellent.

Pour la reproductibilité du score mitotique (variable qualitative ordinale), le kappa de Cohen pondéré de façon linéaire entre les pathologistes lecteurs et entre chaque pathologiste lecteur et le consensus d'expert a été utilisé. Les recommandations de McHugh de 2012 (26), ont été considérées pour l'interprétation de l'accord : en dessous de 0,20 : nul ; entre 0,21 et 0,39 : minimal ; entre 0,40 et 0,59 : faible ; entre 0,60 et 0,79 : modéré ; entre 0,80 et 0,90 : fort ; supérieur à 0,90 : presque parfait.

L'échelle de Koo et Li ainsi que celle de McHugh, plus strictes dans leur définition ont été préférées à l'échelle de Landis et Koch (27), décrite initialement pour l'interprétation du kappa ou celle de Cicchetti (28), décrite pour l'interprétation du coefficient de correlation intraclasse.

Les CCI, kappa de Cohen ont été calculés à l'aide du package "irr" sur les logiciels R (version 4.0.4, cran, The R Foundation for Statistical Computing, http://cran.r-project.org) et R Studio (version 2022.02.2 Build 485 © 2009-2022 RStudio, PBC, http://www.rstudio.com)

Les diagrammes de Bland Altman ont été réalisés sur Excel.

# Résultats

## Reproductibilité entre les pathologistes experts et établissement du consensus

Le consensus pour le score mitotique (considérant l'accord des trois pathologistes en aveugle des deux autres) a été obtenu en première lecture pour 33 cas. Dix-sept cas ont fait l'objet d'une relecture secondaire des trois pathologistes lors d'une séance de relecture commune pour obtenir le consensus sur le score mitotique.

Le **tableau 2** montre le kappa de Cohen pondéré linéairement pour la reproductibilité du score mitotique et le coefficient de corrélation intraclasse pour la reproductibilité du nombre de mitoses pour chaque couple d'experts lors de la détermination du consensus.

Le kappa de Fleiss pour la concordance globale du score mitotique par les experts était de 0,559. Le coefficient de corrélation intraclasse pour la concordance globale du nombre de mitoses par les experts était de 0,804 (IC à 95 % 0,710 - 0,876).

### Reproductibilité inter-observateur des pathologistes lecteurs avec et sans l'algorithme

Pour l'évaluation du compte mitotique sur les 50 cas de carcinome infiltrant, le coefficient de corrélation intraclasse était de 0,591 (IC à 95% 0,375 – 0,746) sans l'utilisation de l'algorithme et de 0,883 (IC à 95% 0,803 – 0,932) avec l'utilisation de l'algorithme.

Pour l'évaluation du score mitotique correspondant, le Kappa de Cohen pondéré de façon linéaire était de 0,482 (IC à 95% 0,231 – 0,733) sans l'utilisation de l'algorithme et de 0,672 (IC à 95% 0,461 – 0,882) avec utilisation de l'algorithme. Le **tableau 3** montre l'accord inter-observateur pour les scores mitotiques entre les pathologistes 1 et 2 avec et sans l'utilisation de l'algorithme. La **figure 3** montre les diagrammes de Bland-Altman avec la différence moyenne et les limites d'accord à 95 % pour la comparaison des comptes mitotiques entre les deux pathologistes lecteurs avec et sans l'utilisation de l'algorithme.

### Reproductibilité inter-observateur entre pathologistes lecteurs et consensus d'expert avec et sans l'algorithme

Le **tableau 4** montre l'accord inter-observateur pour les scores mitotiques entre chaque pathologiste lecteur et le consensus d'experts avec et sans algorithme. Pour l'évaluation du score mitotique sur les 50 cas de carcinome infiltrant, entre le pathologiste 1 et le consensus d'expert, le kappa de cohen pondéré de façon linéaire était de 0,378 (IC à 95% 0,179 – 0,577) sans l'utilisation de l'algorithme et de 0,629 (IC à 95% 0,437 – 0,820) avec utilisation de l'algorithme.

Entre le pathologiste 2 et le consensus d'expert, le kappa de cohen pondéré de façon linéaire était de 0,457 (IC à 95% 0,267 – 0,647) sans l'utilisation de l'algorithme et de 0,726 (IC à 95% 0,575 – 0,876) avec utilisation de l'algorithme.

## Discussion

Bien que le score mitotique inclus dans le grade d'Elston-Ellis soit un critère pronostic majeur du cancer du sein, son évaluation est sujette à une importante variabilité intra et inter-observateur, tant pour la sélection de la zone de comptage que pour le comptage lui-même

(5). Des études récentes ont montré que l'utilisation des lames numérisées pouvait entraîner une sous-estimation du score mitotique de 20% (12). Avec le développement de la pathologie numérique, de plus en plus de services de pathologie seront confrontés à la problématique de l'évaluation des mitoses sur lame numérique.

Dans cette étude, nous avons exploré l'impact de l'utilisation d'un algorithme de détection de mitoses par le pathologiste sur lame entière numérisées.

Nous avons montré que l'utilisation d'un algorithme permettant d'identifier et de localiser les mitoses sur l'ensemble de la surface tumorale analysée puis de proposer automatiquement des zones de *hotspot* permettait d'améliorer la reproductibilité inter-observateur entre pathologistes lecteurs. Dans notre étude, la reproductibilité des deux pathologistes lecteurs sur lame entière était améliorée tant sur le compte mitotique ou l'accord passait de moyen à bon selon l'échelle de Koo et Li que sur le score correspondant où l'accord passait de faible à modéré selon l'échelle de McHugh.

Pour les deux pathologistes lecteurs, l'utilisation de l'algorithme permettait également d'améliorer la reproductibilité avec le consensus d'experts et donc l'exactitude du score établi. L'accord avec le consensus passait de minimal à modéré pour le pathologiste 1 et il passait de faible à modéré pour le pathologiste 2.

Ces niveaux de concordances obtenus lors de la lecture des cas avec l'algorithme étaient comparables à ceux obtenus entre chaque couple d'experts lors de la détermination du consensus. Pour les deux pathologistes lecteurs, une réduction du nombre de discordances de score 1 vs 3 par rapport au consensus a été constatée.

Plusieurs études ont déjà montré que l'utilisation d'une zone de *hotspot* prédéfinie permettait d'améliorer la reproductibilité inter-observateur du compte mitotique. Dans l'étude de Balkenhol *et al.* portant sur 90 cas de cancer du sein, la reproductibilité de deux pathologistes dans l'évaluation du score mitotique était améliorée sur zone de *hotspot* de 2mm² proposée par algorithme sur lame numérique par rapport à la lame de verre (21). Le kappa pondéré de façon linéaire portant sur l'évaluation des cas sur lame de verre était de 0,689 (IC à 95% 0,580–0,799) et de 0,814 (IC à 95% 0,719 – 0,909) lors de l'évaluation sur zone de *hotspot* de 2mm² proposée par l'algorithme. Ce résultat était plus ou moins attendu dans la mesure où dans cette situation la seconde lecture contraint la zone de comptage par rapport à la lame de verre où les pathologistes étaient libres de faire leur compte où ils le souhaitaient. De plus dans l'étude seule la zone de *hotspot* était visible par les pathologistes et les mitoses détectées par l'algorithme n'étaient pas visibles par les pathologistes lors de la lecture des cas ce qui diffère de notre méthodologie.

Dans ces études, les pathologistes comptaient systématiquement dans le même *hotspot* proposé par l'algorithme ce qui constitue un biais expliquant en partie l'amélioration de la reproductibilité.

L'étude de balkenhol *et al.* suggère par ailleurs que le passage des lames de verre au lames numérisées semble introduire moins de variabilité que la sélection des *hotspots* ou la variabilité causée par la présence de différents observateurs. Le passage à une évaluation entièrement automatisée de la densité mitotique sur lame numérique peut potentiellement réduire à la fois la variabilité de l'observateur et la variabilité causée par la sélection des *hotspots*

Dans notre étude, même lors de l'utilisation de l'algorithme, les pathologistes étaient libres de compter les mitoses dans la zone qu'ils estimaient la plus appropriée et restaient libres d'ignorer les recommandations de l'algorithme s'ils estimaient que les données présentées n'étaient pas pertinentes. Cette démarche s'inscrit dans une logique d'utilisation en condition de routine où l'algorithme se présente comme un assistant au comptage librement intégré à la pratique de chaque pathologiste.

A notre connaissance il s'agit de la première étude évaluant la reproductibilité inter-observateur du score mitotique des pathologistes sur lame entière avec et sans algorithme lors de laquelle les pathologistes restent libres de choisir la zone de comptage à chacune des phases de lectures.

Une des forces de notre étude est que la sélection des cas s'est faite au fur et à mesure dans le cadre du flux de travail du service de pathologie de Bicêtre, ceci dans une logique d'utilisation future de l'algorithme en routine. De plus les pathologistes lecteurs étaient de niveau d'expertise différent. Dans le cadre de l'évaluation de la reproductibilité du compte mitotique c'est également la première étude a avoir mis en place un consensus d'expert pour déterminer la vérité de terrain.

Une des limites de notre étude est que nous n'avons pas réalisé d'évaluation de la reproductibilité du compte et du score mitotique par rapport à la lame de verre. Pour évaluer l'apport de l'algorithme en termes de reproductibilité qui était le critère de jugement principal, nous avons considéré qu'il était plus pertinent d'avoir le même support de lecture pour tous les pathologistes à toutes les phases de l'étude. Cependant la lame de verre reste pour l'heure le gold standard pour l'évaluation des mitoses dans le cadre du grade selon Elston-Ellis.

A l'avenir il pourrait être intéressant de comparer la reproductibilité des pathologistes entre une lecture sur lame de verre et une lecture sur lame numérique accompagnée de l'algorithme afin de voir si son utilisation permet de compenser la sous-estimation du compte constatée sur lame numérique par l'équipe de Nottingham.

Dans l'étude de Balkenhol *et al.*, les auteurs avaient également noté que dans le cas des tumeurs comportant peu de mitoses, l'algorithme avait défini la zone de *hotspot* à partir de détections erronées de faux-positifs, soulignant par là-même l'importance de permettre au pathologiste d'avoir un visuel sur les figures proposées par l'algorithme pour définir la zone de comptage. Ce regard critique est indispensable. En effet, dans le cadre de notre étude, certaines mitoses authentiques étaient également proposées par l'algorithme avec un indice de confiance inférieur à certains faux-positifs.

A notre connaissance, ce travail est le premier à être concrètement engagé vers l'intégration d'un algorithme de détection de mitoses au flux de travail de routine du pathologiste.

Cet algorithme n'a pas vocation à se substituer à l'avis du pathologiste. Il ne peut faire que des propositions d'images que le pathologiste choisit de considérer ou non comme mitose.

Une des limites actuelles des algorithmes d'intelligence artificielle dans la reconnaissance des mitoses est leur manque de capacité à considérer les informations contextuelles. Ces données s'intègrent en effet de façon intuitive dans l'esprit du pathologiste lorsqu'il analyse ses lames et en pratique, deux cellules d'aspect comparable pourraient être considérées en mitoses ou en "non-mitoses" selon l'aspect global de la tumeur, des cellules adjacentes, de la présence de remaniements nécrotiques ou d'artefacts. De nombreuses études de validation sur des bases de données différentes seront nécessaires avant d'envisager la possibilité de déléguer l'établissement du score mitotique à un algorithme.

D'autres questions restent en suspens. Comment ce type d'outil serait-il accepté par la communauté des pathologistes? Comment serait-il concrètement intégré dans le flux de travail de nos laboratoires? Les analyses seraient-elles faites avant même que le pathologiste accède à la lame numérique par prescription automatique lors de l'enregistrement du cas dans le système de gestion de laboratoire? Où bien l'outil serait-il intégré comme une nouvelle fonctionnalité du système de gestion d'image utilisé, appliqué à la demande du pathologiste lors de la lecture des lames du cas? Comment le pathologiste s'appropriera-t-il les données présentées par l'algorithme et comment les intégrera-t-il dans son compte rendu? Quels seraient les contrôles qualité à mettre en place pour s'assurer que les performances de l'algorithme restent constantes? L'évaluation médico-économique de l'outil devra également être considérée.

Au-delà des problématiques de reproductibilité, l'utilité pratique de l'outil développé dans notre étude devra être prise en compte. L'apport de celui-ci en terme de gain de temps, d'ergonomie et de praticité pour le compte mitotique sur lame numérique devra faire l'objet d'études spécifiques.

Les résultats de notre étude et les questions évoquées pourraient s'intégrer dans une réflexion plus globale qui pourrait amener à modifier la méthode d'évaluation du score mitotique. Le nombre de mitoses requis par unité de surface pour définir chacun des trois scores pourrait également être amené à évoluer pour tenir compte des différences observables sur lame numérique et des modifications éventuelles induites par l'utilisation des outils d'intelligence artificielle. L'évolution des pratiques qui pourrait en découler nécessiterait des études supplémentaires à plus grande échelle et intégrant la survie des patientes pour valider la valeur pronostique de ces nouvelles méthodes d'analyse.

En conclusion, nous avons montré que l'utilisation de l'algorithme développé constituait une approche viable pour assister le pathologiste lors de l'évaluation du score mitotique des carcinomes mammaires infiltrants sur lame entière numérisée. Son utilisation permettant d'améliorer la reproductibilité inter-observateur entre pathologistes lecteurs et l'exactitude du score établi par consensus d'expert.

L'utilisation d'un tel outil constitue une nouvelle approche dans l'évaluation du score mitotique qui pourrait conduire à une évolution des pratiques. D'autres études seront nécessaires pour en définir la place et les modalités d'utilisations concrètes.

# Tableaux et figures

**Tableau 1 :** Caractéristiques des patients et des tumeurs de la cohorte utilisée pour l'étude

|  | Cohorte ($n$ = 50)<br>Nombre de cas (%) |
|---|---|
| Genre |  |
| Femme | 50 (100.0) |
| Age |  |
| > ou égal à 50 ans | 42 (84.0) |
| < 50 ans | 8 (16.0) |
| stade pathologique de la tumeur (pour les résections mammaires uniquement - 25 cas)* |  |
| pT1 | 18 (72.0) |
| pT2 | 4 (16.0) |
| pT3 | 1 (4.0) |
| pT4 | 2 (8.0) |
| stade pathologique des ganglions lymphatiques (pour les résections mammaires uniquement - 25 cas)* |  |
| N0 (incluant les cellules tumorales isolées) | 12 (48.0) |
| N1 | 9 (36.0) |
| N2 | 0 (0.0) |
| N3 | 1 (4.0) |
| Nx | 3 (12.0) |
| Sous type histologique |  |
| Carcinome infiltrant de type non spécifique | 39 (78.0) |
| Carcinome infiltrant de type non spécifique avec différenciation neuroendocrine | 2 (4.0) |
| Carcinome infiltrant mixte de type non spécifique et mucineux | 1 (2.0) |
| Carcinome infiltrant mixte de type non spécifique et micro-papillaire | 1 (2.0) |
| Carcinome lobulaire infiltrant | 6 (12.0) |
| Carcinome micropapillaire | 1 (2.0) |
| Statut ER/PR, HER2 |  |
| ER+/PR+/HER2- | 39 (78.0) |
| ER+/PR-/HER2- | 6 (12.0) |
| ER-/PR-/HER2- | 2 (4.0) |
| ER-/PR-/HER2+ | 3 (6.0) |
| Emboles lymphatiques |  |
| Négatifs | 47 (94.0) |
| Positifs | 3 (6.0) |
| Carcinome *in-situ* associé |  |
| Oui | 18 (36.0) |
| Non | 32 (64.0) |

*Pour les 25 cas de résection mammaire, le stade de la tumeur primaire et le stade des ganglions lymphatiques régionaux ont été classés selon la classification TNM 2017 des tumeurs malignes (UICC - 8e édition).

**Figure 1 :** pipeline général de l'algorithme utilisé pour l'étude.

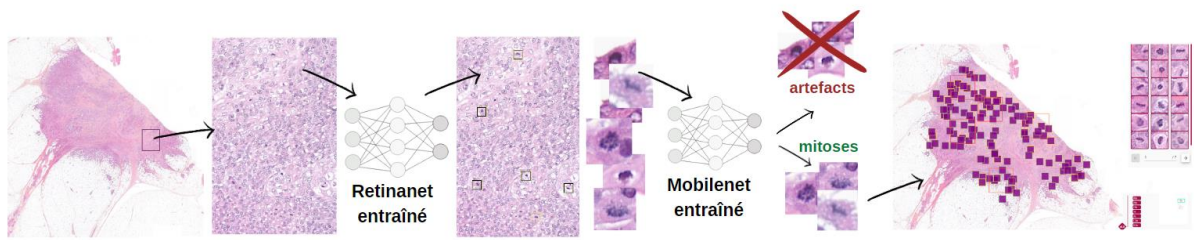

Illustration – Primaa

**Figure 2 :** présentation de l'interface de visualisation (Cleo) ; détection des mitoses (petits carrés) ; zones de plus forte densité mitotique proposées par l'algorithme (grands carrés) ; visualisation des mitoses détectées (onglet latéral).

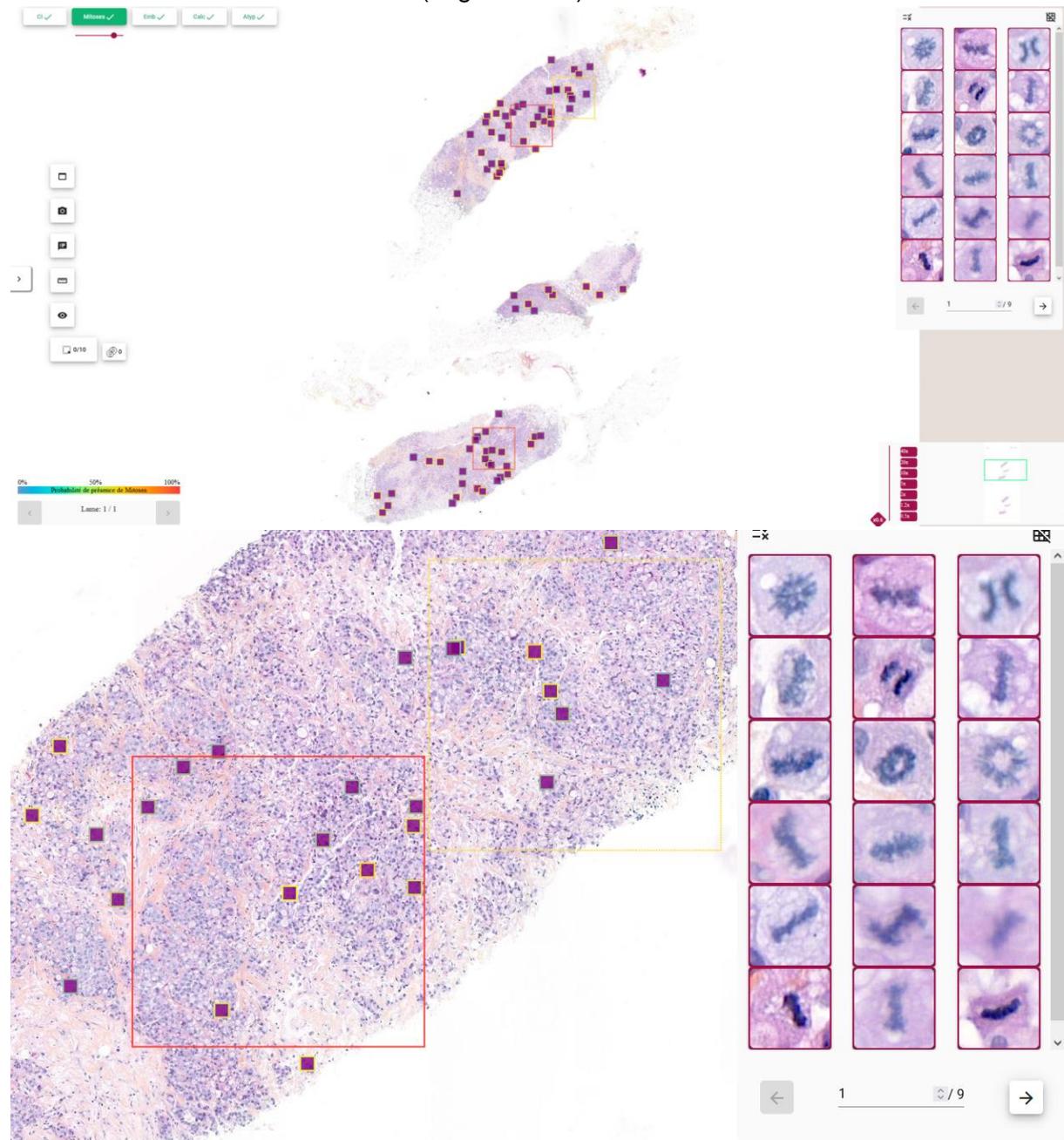

Illustrations – Primaa

**Tableau 2** : comparaisons croisées, kappa de Cohen linéaire pondéré pour la reproductibilité du score mitotique et coefficient de corrélation intraclasse pour la reproductibilité du nombre de mitoses pour chaque couple d'experts lors de la détermination du consensus

| | | Expert 1 | | | |
|---|---|---|---|---|---|
| | Score | 1 | 2 | 3 | Total |
| Expert 2 | 1 | 24 | 1 | 1 | 26 |
| | 2 | 7 | 3 | 1 | 11 |
| | 3 | 1 | 2 | 10 | 13 |
| | Total | 32 | 6 | 12 | 50 |

Comparaison croisée de l'accord inter-observateur pour les scores mitotiques entre les experts 1 et 2. Le Kappa de Cohen pondéré linéairement était de 0,655 (IC 95 % 0,479 - 0,831).
Coefficient de corrélation intraclasse pour le nombre de mitoses : 0,786 (IC 95 % 0,651 - 0,872)

| | | Expert 1 | | | |
|---|---|---|---|---|---|
| | Score | 1 | 2 | 3 | Total |
| Expert 3 | 1 | 29 | 3 | 2 | 34 |
| | 2 | 2 | 2 | 1 | 5 |
| | 3 | 1 | 1 | 9 | 11 |
| | Total | 32 | 6 | 12 | 50 |

Comparaison croisée de l'accord inter-observateur pour les scores mitotiques entre les experts 1 et 3. Le Kappa de Cohen pondéré linéairement était de 0,677 (IC 95 % 0,482 - 0,871).
Coefficient de corrélation intraclasse pour le nombre de mitoses : 0,761 (IC 95 % 0,614 - 0,857)

| | | Expert 2 | | | |
|---|---|---|---|---|---|
| | Score | 1 | 2 | 3 | Total |
| Expert 3 | 1 | 27 | 2 | 0 | 26 |
| | 2 | 7 | 1 | 1 | 11 |
| | 3 | 4 | 5 | 3 | 13 |
| | Total | 34 | 5 | 11 | 50 |

Comparaison croisée de l'accord inter-observateur pour les scores mitotiques entre les experts 2 et 3. Le Kappa de Cohen pondéré linéairement était de 0,627 (IC 95 % 0,434 - 0,820).
Coefficient de corrélation intraclasse pour le nombre de mitoses : 0,851 (IC 95 % 0,752 - 0,912)

Kappa de Fleiss pour la concordance globale des experts sur le score mitotique : 0,559
Coefficient de corrélation intraclasse pour la concordance globale des experts sur le nombre de mitoses : 0,804 (IC 95 % 0,710 - 0,876)

**Tableau 3 :** comparaisons croisées de l'accord inter-observateur pour les scores mitotiques entre les pathologistes 1 et 2 avec et sans algorithme.

| | | Pathologiste 1 sans l'algorithme | | | |
|---|---|---|---|---|---|
| | Score | 1 | 2 | 3 | Total |
| Pathologiste 2 sans l'algorithme | 1 | 34 | 4 | 1 | 39 |
| | 2 | 4 | 2 | 1 | 7 |
| | 3 | 0 | 2 | 2 | 4 |
| | Total | 38 | 8 | 4 | 50 |
| Comparaison croisée de l'accord inter-observateur pour les scores mitotiques entre les pathologistes 1 et 2 sans algorithme Le Kappa de Cohen pondéré linéairement était de 0,482 (IC 95 % 0,231 - 0,733) sans l'utilisation de l'algorithme. Coefficient de corrélation intraclasse pour le nombre de mitoses sans algorithme : 0,591 (IC 95 % 0,375 - 0,746) | | | | | |
| | | Pathologiste 1 avec l'algorithme | | | |
| | Score | 1 | 2 | 3 | Total |
| Pathologiste 2 avec l'algorithme | 1 | 30 | 4 | 1 | 35 |
| | 2 | 1 | 4 | 1 | 6 |
| | 3 | 2 | 0 | 7 | 9 |
| | Total | 33 | 8 | 9 | 50 |
| Comparaison croisée de l'accord inter-observateur pour les scores mitotiques entre les pathologistes 1 et 2 avec l'algorithme. Le Kappa de Cohen pondéré linéairement était de 0,672 (IC 95 % 0,461 - 0,882) avec l'utilisation de l'algorithme. Coefficient de corrélation intraclasse pour le nombre de mitoses sans algorithme : 0,883 (IC 95 % 0,803 - 0,932). | | | | | |

**Tableau 4** : comparaisons croisées de l'accord inter-observateur pour les scores mitotiques entre chaque lecteur pathologiste et le consensus d'experts avec et sans algorithme

|  |  | Pathologiste 1 sans l'algorithme | | | |
|---|---|---|---|---|---|
| Consensus d'experts | Score | 1 | 2 | 3 | Total |
|  | 1 | 27 | 2 | 0 | 29 |
|  | 2 | 7 | 1 | 1 | 9 |
|  | 3 | 4 | 5 | 3 | 12 |
|  | Total | 38 | 8 | 4 | 50 |

Comparaison croisée de l'accord inter-observateur pour les scores mitotiques entre le pathologiste 1 sans algorithme et le consensus d'experts
Le Kappa de Cohen pondéré linéairement était de 0,378 (IC 95 % 0,179 - 0,577) sans l'utilisation de l'algorithme.

|  |  | Pathologiste 1 avec l'algorithme | | | |
|---|---|---|---|---|---|
| Consensus d'experts | Score | 1 | 2 | 3 | Total |
|  | 1 | 27 | 1 | 1 | 29 |
|  | 2 | 4 | 4 | 1 | 9 |
|  | 3 | 2 | 3 | 7 | 12 |
|  | Total | 33 | 8 | 9 | 50 |

Comparaison croisée de l'accord inter-observateur pour les scores mitotiques entre le pathologiste 1 avec l'algorithme et le consensus d'experts
Le Kappa de Cohen pondéré linéairement était de 0,629 (IC 95 % 0,437 - 0,820) avec l'utilisation de l'algorithme.

|  |  | Pathologiste 2 sans l'algorithme | | | |
|---|---|---|---|---|---|
| Consensus d'experts | Score | 1 | 2 | 3 | Total |
|  | 1 | 28 | 1 | 0 | 29 |
|  | 2 | 9 | 0 | 0 | 9 |
|  | 3 | 2 | 6 | 4 | 12 |
|  | Total | 39 | 7 | 4 | 50 |

Comparaison croisée de l'accord inter-observateur pour les scores mitotiques entre le pathologiste 2 sans algorithme et le consensus d'experts
Le Kappa de Cohen pondéré linéairement était de 0,457 (IC 95 % 0,267 - 0,647) sans l'utilisation de l'algorithme.

|  |  | Pathologiste 2 avec l'algorithme | | | |
|---|---|---|---|---|---|
| Consensus d'experts | Score | 1 | 2 | 3 | Total |
|  | 1 | 28 | 1 | 0 | 29 |
|  | 2 | 7 | 2 | 0 | 9 |
|  | 3 | 0 | 3 | 9 | 12 |
|  | Total | 35 | 6 | 9 | 50 |

Comparaison croisée de l'accord inter-observateur pour les scores mitotiques entre le pathologiste 2 avec l'algorithme et le consensus d'experts
Le Kappa de Cohen pondéré linéairement était de 0,726 (IC 95 % 0,575 - 0,876) avec l'utilisation de l'algorithme.

**Figure 3 :** diagrammes de Bland-Altman avec différence moyenne et limites d'accord à 95 % (lignes pointillés rouges) pour la comparaison des comptes mitotiques entre les deux pathologistes lecteurs avec et sans l'utilisation de l'algorithme.

A : pathologiste lecteur 1 versus pathologiste lecteur 2 sans utilisation de l'algorithme
B : pathologiste lecteur 1 versus pathologiste lecteur 2 avec utilisation de l'algorithme

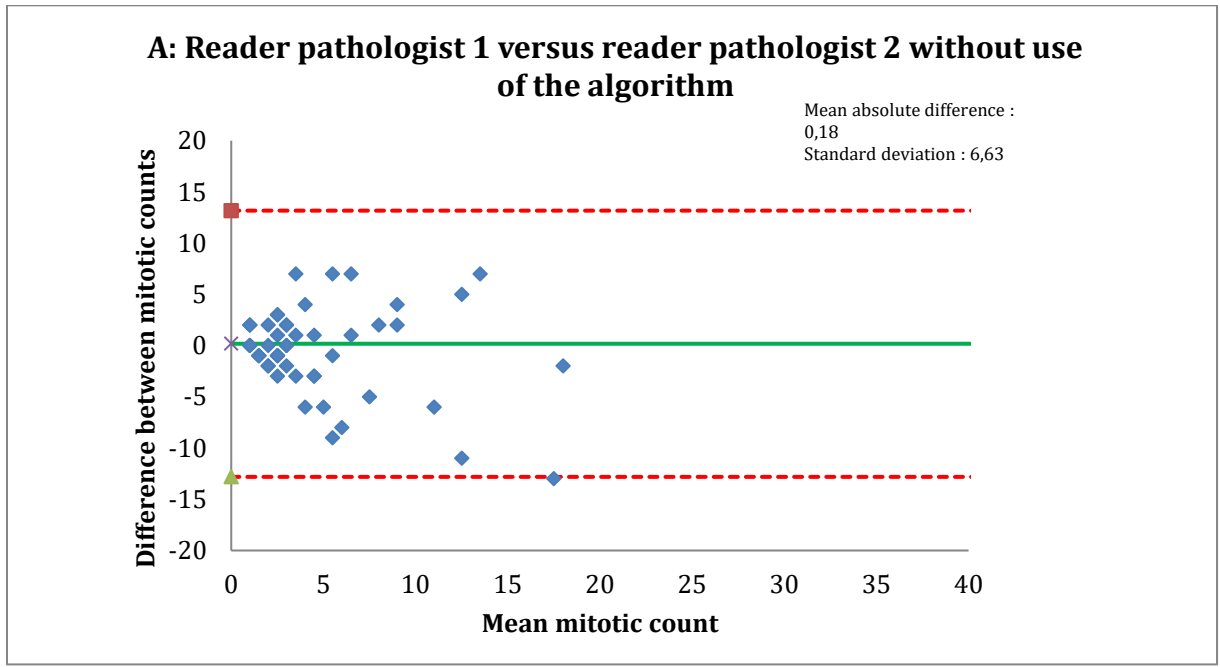

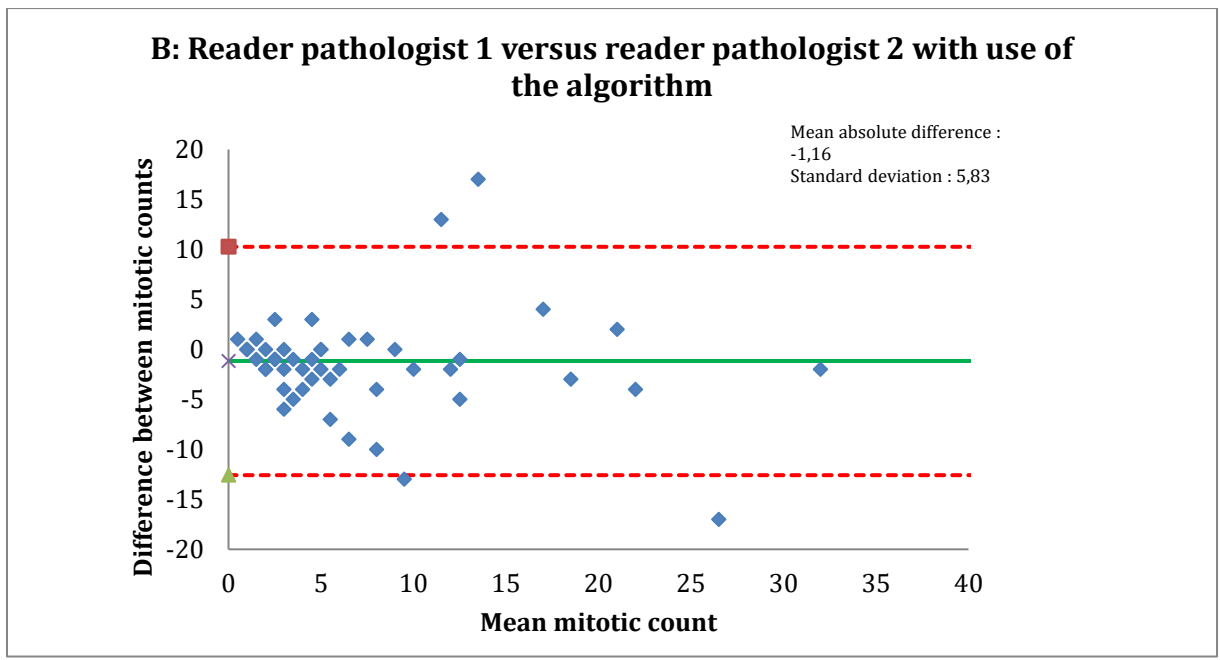

# Bibliographie

# Résumé

**Introduction :**
Le grade de Nottingham est un facteur pronostique majeur du carcinome mammaire infiltrant. Sa détermination nécessite l'évaluation du score mitotique qui est sujet à une faible reproductibilité intra et inter-observateur. Le score mitotique doit être réalisé dans la zone la plus proliférative de la tumeur ce qui est difficile à déterminer mais essentiel. Des outils basés sur l'intelligence artificielle pourraient aider les pathologistes à détecter les mitoses sur lames entières numérisées.

**Objectif de l'étude**
L'objectif de cette étude était d'évaluer l'apport d'un algorithme de détection des mitoses spécifiquement développé pour aider le pathologiste lors de l'évaluation du score mitotique sur lame entière numérisées.

**Méthodes :**
La détection algorithmique des mitoses est un processus en deux étapes : l'algorithme détecte d'abord les objets candidats ressemblant à des mitoses, puis la sélection est affinée par un classificateur. Les régions mitotiques les plus denses sont montrées au pathologiste, qui peut alors établir le score mitotique à l'aide des résultats de l'algorithme. Pour cette étude, trois pathologistes experts ont déterminé le score mitotique de façon consensuelle sur cinquante lames entières numérisées de carcinome mammaire infiltrant. Ces lames ont également été soumises à deux pathologistes lecteurs qui ont évalué le score mitotique de chaque lame deux fois, avec et sans l'aide de l'algorithme, avec une période de *wash-out* de quatre semaines. La reproductibilité inter-observateur a été mesurée en évaluant les scores obtenus avec et sans l'aide de l'algorithme, pour les deux pathologistes lecteurs et a également été mesurée entre chaque pathologiste lecteur et le consensus d'expert afin de déterminer la précision du score établi.

**Résultats :**
Le Kappa de Cohen pondéré linéairement pour l'accord inter-observateur du score mitotique entre les deux pathologistes lecteurs était de 0,482. En utilisant la détection des mitoses générée par l'algorithme, il était de 0,672. Le Kappa de Cohen pondéré linéairement pour la concordance inter-observateur du score mitotique entre chaque pathologiste lecteur et le consensus d'experts était de 0,378 et 0,457 pour les pathologistes 1 et 2 respectivement. En utilisant la détection des mitoses générée par l'algorithme, le kappa était augmenté respectivement à 0,629 et 0,726.

**Conclusion :**
L'utilisation de l'algorithme développé constitue une approche viable pour assister le pathologiste dans l'évaluation du score mitotique du carcinome mammaire infiltrant sur lame entière numérique. Son utilisation permet d'améliorer la reproductibilité inter-observateur entre pathologistes et la précision du score établi par consensus d'experts. L'utilisation d'un tel outil constitue une nouvelle approche dans l'évaluation du score mitotique qui pourrait conduire à une évolution des pratiques.

# Abstract

**Introduction**:
Nottingham grading system is a major prognostic factor for invasive breast carcinoma (IBC). Its determination requires the evaluation of the mitotic score (MS) which is subject to low intra and interobserver reproducibility. The MS shall be performed in the most proliferative area of the tumor, which determination is hard but critical. Artificial intelligence based tools could help pathologists to detect mitosis on whole slide images (WSI).

**Objective**
The aim of this study was to evaluate the contribution of a mitosis detection algorithm specifically developed to assist the pathologist during the evaluation of the MS on WSI.

**Methods:**
Algorithmic mitosis detection is a two-step process : first the algorithm detects candidate objects resembling mitosis, then the selection is refined by a classifier. The densest mitotic regions are shown to the pathologist, then he can establish the MS with algorithm results. For this study, three expert pathologists have determined a consensual ground truth for MS on fifty WSI of IBC. Those slides were also submitted to two readers pathologists who evaluated the MS of each slide twice, with and without the assistance of the algorithm, with a four week wash-out period. Interobserver reproducibility was measured by evaluating the scores obtained with, and without assistance between two readers pathologists and was also measured between each reader pathologist and the expert ground truth to determine the accuracy of the established score.

**Results:**
Baseline linearly weighted Cohen's Kappa for interobserver agreement of MS between two readers pathologists was 0.482. Using the algorithm generated mitotic detection in WSI, the agreement score increased to 0.672. Baseline linearly weighted Cohen's Kappa for interobserver agreement of MS between each reader pathologist and expert consensus was 0.378 and 0.457 for pathologist 1 and 2 respectively. Using the algorithm generated mitotic detection in WSI, the agreement score increased respectively to 0.629 and 0.726.

**Conclusion:**
The use of the developed algorithm constitutes a viable approach to assist the pathologist for the evaluation of the MS of IBC on WSI. Its use makes it possible to improve interobserver reproducibility between pathologists and the accuracy of the score established by expert consensus. The use of such a tool constitutes a new approach in the evaluation of the mitotic score which could lead to an evolution of practices.